\documentclass[twocolumn,showpacs,preprintnumbers,amsmath,amssymb]{revtex4}

\newcommand{\s}{\sigma}
\newcommand{\bs}{\bar\sigma}
\newcommand{\ra}{\rangle}
\newcommand{\la}{\langle}

\usepackage{graphicx}
\usepackage{bm}
\usepackage{amsmath}
\usepackage{amssymb}
\usepackage{color}

\begin{document}

\title{Critical hybridization for the Kondo resonance in gapless systems}

\author{ R. G. Dias}
\author{L\'{\i}dia del Rio}
\author{A. V. Goltsev}
\affiliation{Departamento de F\'{\i}sica, Universidade de Aveiro, \\
3810 Aveiro, Portugal}%

\date{\today}

\begin{abstract}
We study the Kondo resonance in a spin-1/2 single impurity Anderson
model with a gapless conduction band
using the equation of motion approach in order to obtain the impurity spectral function.
We study two different scenarios  for gapless systems: a purely
power-law energy dependence of the density of states or a constant density of states with a
gapless behavior near the Fermi level. We demonstrate that strong electron-electron correlations
lead to a sharp peak  in the impurity spectral function in the case of a
large exchange coupling ($J>J_{c}$) or equivalently,   a strong hybridization ($V>V_{c}$).
This  Kondo-like peak  emerges   much below the Fermi level in the case of a
strongly  depleted  density of states. These results are compared with the ones from renormalization group approaches.
\end{abstract}


\maketitle
\section{Introduction}
The advent of graphene has renewed the interest in the Kondo effect in gapless systems.\cite{Neto2009}
It is known that a strong depletion of the density of states of a conduction band near the Fermi level modifies the
usual  behavior of magnetic and  nonmagnetic impurities.\cite{Withoff1990,Ingersent1996} Impurities in gapless systems have
been addressed, for instance, in the context of d-wave
superconductors \cite{Polkovnikov2001,Vojta2002a,Zhu2001} and more
recently graphene,\cite{Sengupta2008} using a variety of methods,
from renormalization group approaches \cite{Bulla1997,Fritz2005,
Gonzalez-Buxton1998,GonzalezBuxton1996}  to the large N analysis
\cite{Polkovnikov2002} and T-matrix calculations.\cite{Balatsky2006}
The most relevant modification  in gapless systems of the usual Kondo behavior
is the requirement of a finite critical Kondo coupling strength  in
order to observe the Kondo resonance.\cite{Withoff1990,Sengupta2008}
While several renormalization group studies have addressed this point \cite{Withoff1990,Sengupta2008,Ingersent1996},
an equivalent study from the point of view of the
impurity spectral function in the case of the single impurity Anderson model
has not been carried out as far as we know. In the present manuscript, we address this problem
using the equation of motion (EOM) approach  which has been   successful in describing
the Kondo resonance in systems with a finite density of states at the Fermi level.\cite{LACROIX1981}
The nonmagnetic case when  the on-site Coulomb repulsion is zero (\(U=0\)), and the strong
on-site Coulomb repulsion limit \((U\rightarrow \infty) \)  of the gapless single impurity Anderson model are considered.
Note that  above the Kondo temperature, the  $U\rightarrow \infty$  spectral function should recover the $U=0$ features.
We will show that even for $U=0$, an additional peak may appear in the spectral function due to the
depletion of the density of states which should not be confused with a Kondo peak. We will demonstrate that the critical exchange coupling depends strongly
on the density of states profile in the gapless region. Furthermore, the strong electron-electron
correlations can produce a strong Kondo-like peak in the impurity spectral function even much
below the Fermi level in the depletion region. This Kondo-like peak emerges if the hybridization between the conduction
band
and the impurity is larger than a critical value, which in turn is determined by the energy dependence of the density of states in the depletion region,   and grows with decreasing temperature.
\section{Gapless Anderson Hamiltonian}
\begin{figure}[tp]
      \includegraphics[width=0.45\textwidth,clip]{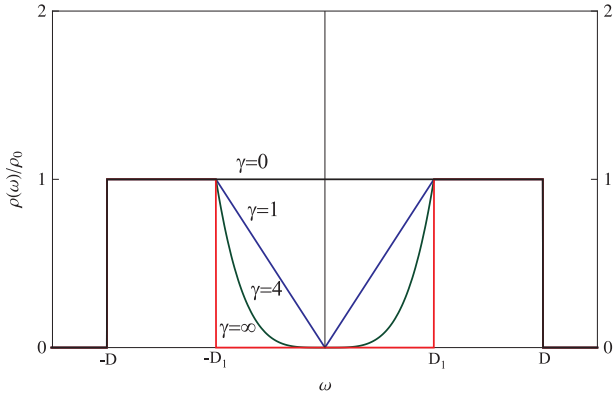}
      \caption{\label{fig:densityofstates}
        The density of states $\rho(\omega)$ of the conduction band
        used in this article. The density of states is constant in the energy range
        $D_1 <\vert \omega \vert <D$ and has a power-law dependence
        $\rho(\omega)=\alpha \vert \omega \vert^\gamma$ for $\vert \omega \vert <D_1$.
        The  density of states  profile for several $\gamma$ values is displayed.
        Note that $\gamma = \infty$ implies the existence of a $2D_1$  energy gap.}
\end{figure}

The single impurity Anderson model is given by
\begin{eqnarray}
        H &=&\sum_{k\s}\varepsilon_{k}\hat{n}_{k\s} + \sum_{\s}
        E_0\hat{n}_{d\s}\nonumber\\
        &&+ \frac12U\sum_{\s}\hat{n}_{d\s}\hat{n}_{d\bs} +
        V\sum_{k\s}(c_{k\s}^{+}d_{\s}+d_{\s}^{+} c_{k\s}). \label{eq:hamiltonian}
\end{eqnarray}
where the standard notations\cite{Hewson1993} are used: \(V\) is the hybridization between conduction and impurity
states, and \(E_{0}\) is the impurity energy level.
In order to describe  gapless systems, a density of states with a
power-law energy dependence
is usually considered  in a certain energy interval around
the Fermi energy which we set at $\varepsilon_F=0$. We will assume that the density of states of the conduction band is constant outside
this interval, $ \rho(\varepsilon)=\rho_{0}$, for
$D_{1}<\vert \varepsilon \vert<D,$ and
$\rho(\varepsilon)= \rho_1(\varepsilon)=\alpha \vert \varepsilon \vert^\gamma$ for $\vert \varepsilon \vert<D_{1}$
with $\alpha=\rho_{0}/D_{1}^\gamma$. Therefore,
$\rho(\varepsilon)$ is zero at the Fermi level. The bandwidth is \(2D\).  In Fig.~\ref{fig:densityofstates}, the density of states
for several values of $\gamma$ is shown.  In graphene,  \(\gamma=1\). If \(D_{1}=0\), then one has a constant density of states with the finite bandwidth. If \(D_{1}=D\), only the power-law dependence of the
density of states is considered. $\gamma=\infty$  corresponds to a system with a \(2D_{1} \) gap.
Below, we will discuss in detail  the case  $E_0 < -D_1$.
\section{Equation of motion method}
The equation of motion approach is basically the successive application of the
result
$
        \omega G_{AB}(\omega)=\la\{A,B\}\ra+\la\la[A,H];B\ra\ra_\omega
$
to the impurity Green's function and to the other  Green's function
generated in the process.\cite{LACROIX1981} Here we have  adopted the Zubarev notation
for the retarded Green's function,  $G_{A,B}(\omega) = \la\la A; B\ra\ra_{\omega}$.\cite{Zubarev1960}
As usual, it is implicit that \(\omega\rightarrow\omega+i\eta\) where
\(\eta\) is a infinitesimal positive constant. In the case of the Anderson model,
one obtains  the following EOM for the impurity Green's function
\begin{multline}
        \left(\omega-E_{0}-
        \sum_k\frac{V^2}{\omega - \varepsilon_k}\right)  \la\la
        d_{\s};d_{\s}^{+}\ra\ra_\omega =1  \\
        + U\la\la\hat{n}_{d\bs}d_{\s};d_{\s}^{+}\ra\ra_\omega.
        \label{eq:eom}
\end{multline}
If the on-site repulsion between electrons is neglected (\(U=0\)), this equation
becomes a closed EOM. Below, we discuss this \(U=0\) case in detail before addressing the $U
\rightarrow \infty$ limit for two reasons. First, due to the depletion of the density of states
at the Fermi level,
the \(U=0\) impurity spectral function will have a non trivial dependence on the impurity energy  level
and on the hybridization value \(V\). In particular, it may display  more structure than the typical  Lorentzian  profile.
 This more complex profile is the background  against which the correlation effects
 will be observed  in the $U\rightarrow \infty$ limit. Second, the $U=0$ features should be recovered in  the  $U\rightarrow \infty$  spectral function above the Kondo temperature.

\section{Resonant level ($U=0$)}
\begin{figure}[tp]
      \includegraphics[width=0.45\textwidth,clip]{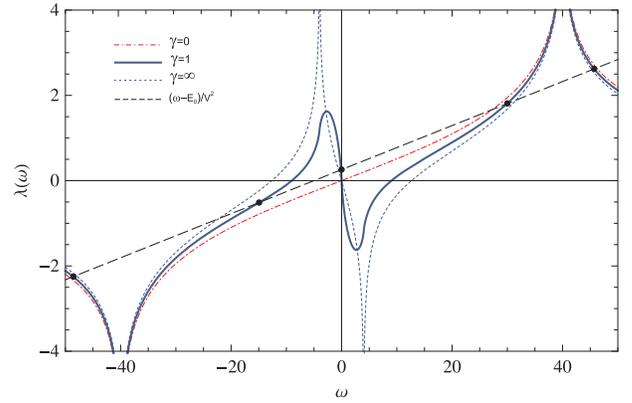}
      \caption{\label{fig:spectralnonint}  The real part $\lambda (\omega)$ of the self-energy
      $\Sigma_0$, Eq.~(\ref{eq:5}),
      for the resonant level (\(U=0\))  exhibits logarithmic divergences
      at the band edges and a non-monotonous behavior in the gapless frequency region.
      The dashed line  represents  the line  $(\omega -E_0)/V^2$. The black points are
      the roots of $\omega -E_0-\Lambda(\omega) $.  For \(\gamma=\infty\) (the gapped spectrum),
      additional logarithmic divergences are present at the gapless region
      edges. Parameters: \(E_{0}=-5,\)
      $\rho_0=1$, $V=4.4$, $D_1=4$ and $D=40$.}
\end{figure}

The  self-energy for  \(U=0\) can be easily obtained for integer \(\gamma\) and  is given by
\begin{equation}
        \Sigma_0=\sum_k\frac{V^2}{\omega - \varepsilon_k} =  \Lambda(\omega)-i \Delta(\omega),
                \label{eq:sigma0}
\end{equation}
with
\begin{eqnarray}
        \frac{\Delta(\omega)}{V^2}
        = \tau(\omega) &=&
        \pi  \rho (\omega)+\eta, \\
        \frac{\Lambda(\omega)}{V^2}
        = \lambda (\omega)&=&
        B(\omega)+P_{n}(\omega) \label{eq:5} \nonumber\\
        &+&
        \rho_0\ln \left\vert \frac{(D+\omega)(D_{1}-\omega)}{(D-\omega)(D_{1}+\omega)}
        \right\vert,
\end{eqnarray}
where
\begin{equation}
        B(\omega)=
        \left\{
        \begin{array}{cc}
                \alpha \omega^\gamma\ln
                \left\vert\dfrac{\omega^2}{D_1^2-\omega^2}
                \right\vert,
                &
                \gamma \text{ odd},
                \\
                & \\
                \alpha \omega^\gamma\ln
                \left\vert \dfrac{D_{1}+\omega}{ D_{1}-\omega}\right\vert,                        &
                \gamma \text{ even},
        \end{array}
        \right.
\end{equation}
and  \(P_{n}(\omega)\) is a polynomial of degree \(n<\gamma\).
For \(\gamma=1\) and \(\gamma=\infty\) (a gapped conduction band),
\(P_{n}(\omega)=0\).  In Fig.~\ref{fig:spectralnonint},
the real part  of the self-energy $\Sigma_0$  is displayed for \(\gamma=0\), \(\gamma=1\) and
\(\gamma=\infty\) (the gapped spectrum).

The spectral function  \(\mathcal{A}( \omega )=-2 \, \text{Im}\, \la\la
        d_{\s};d_{\s}^{+}\ra\ra_ \omega \) for a system with finite bandwidth (gapless or not) is given by
\begin{multline}
        \mathcal{A}( \omega )=\frac{2 \Delta(\omega)}{[\omega -E_0-\Lambda(\omega)]^2+\Delta^2
        (\omega)}         \\
        +2\pi\Theta(D-\vert \omega \vert) \sum_i \vert1-\Lambda'(\omega)\vert^{-1}\delta (\omega -\omega_i)
\label{eq:spectralfunction}
\end{multline}
where $\omega_i$ are the roots of  $\omega -E_0-\Lambda(\omega)=0$.
Note that the Dirac delta points outside the band continuum
play an important role if the hybridization is not small. These points  are essential in order to satisfy
the sum rule of the spectral function. This can be more clearly understood if the continuous term is written as
\begin{equation}
         \dfrac{1}{V^2}
         \frac{2 \gamma(\omega)}{\left[\dfrac{\omega}{V^2}
         -\dfrac{E_0}{V^2}-\lambda(\omega)\right]^2+\tau^2(\omega)}.
         \label{eq:spectralcont}
\end{equation}
Therefore, for large $V$ and a constant density of states most of the spectral weight
is associated with the Dirac delta points outside the band continuum since
the total  spectral weight for \(\vert \omega\vert <D  \) decays as \(1/V^{2}\).
As can be inferred from Fig.~\ref{fig:spectralnonint}, for large $V$ [or equivalently a small slope
of the straight line  $(\omega -E_0)/V^2$], the Dirac delta points are located at large \(\omega\).
For large $\omega$,  $\Lambda(\omega) \sim V^{2}C/\omega
$ where $C \sim \rho_0 D$ and therefore $\omega_i=\pm V\sqrt{C}$
leading to   \(\vert1-\Lambda'(\omega)\vert^{-1}
\sim 1/2\). Thus the sum rule is satisfied. For small \(V\), these Dirac delta points
are located near the boundaries of the band due to the log-like behavior of $\Lambda(\omega)$.
The same conclusion holds for a power-law density of states.

For a constant density of states, $ \lambda(\omega)$ is monotonous for $\vert \omega
\vert < D$ and the minima of the first term in the denominator of the spectral function given
by  Eq.~(\ref{eq:spectralcont}) are typically  roots of $\omega -E_0-\Lambda(\omega) $.
In the case of a power-law density of states,
additional minima appear  due to the non-monotonous behavior
of $\Lambda(\omega)$ within the gapless region. These minima are not necessarily
roots of $\omega -E_0-\Lambda(\omega) $, but lead to an additional structure in the spectral function.
\begin{figure}[tp]
      \includegraphics[width=0.45
      \textwidth,clip]{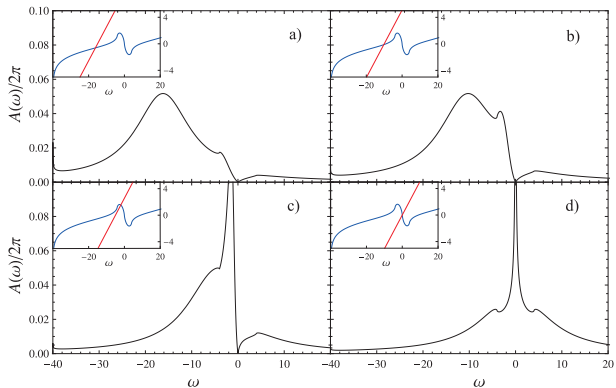}
      \caption{\label{fig:severalE0array}
      The spectral function \(A(\omega)\), from Eq.~(\ref{eq:spectralfunction}), for the resonant level (\(U=0\)) for $ \gamma=1$ and several values of the impurity level energy: a)  \(E_{0}=-15\);
      b)  \(E_{0}=-10\); c)  \(E_{0}=-5\); d)  \(E_{0}=0\).
      Inset: The real part of the self-energy \(\lambda (\omega)\) and the line $(\omega -E_0)/V^2$.
      It is the  non-monotonous  behavior of \(\lambda (\omega)\) that generates additional structure  in the spectral function
      for certain  intervals of the parameters. For small $V$, this structure disappears. Other
      parameters:
      $\rho_0^{1/2} V=1.4$, $D_1=4$ and $D=40$. }
\end{figure}
In Fig.~\ref{fig:severalE0array}, the spectral function for \(U=0\) at several values of $E_{0}$ is displayed.
For general \(\gamma\), one has a $\vert \omega \vert^\gamma $
behavior of the spectral function near the Fermi energy,  except for \(E_{0}=0\) as can be observed
in Fig.~\ref{fig:severalE0array}.
It is worthwhile
to emphasize that the impurity spectral function exhibits the same depletion at the Fermi level as the density of states
of the conduction band.

Note that the value of  $V$ determines the slope of the line $(\omega -E_0)/V^2$
leading to different behavior of the spectral function. A small $V$
implies a large slope and typically only one root of $\omega -E_0-\Lambda(\omega) $
for $\vert \omega \vert <D$. For large $V$, one has an almost horizontal line intersecting
the $\lambda (\omega )$ curve in Fig.~\ref{fig:spectralnonint}.
The same reasoning applies in the interacting case that will be considered below.  Since the \(V^{2}\) terms appear always associated to a \(\rho_0\)
factor in the spectral function, these parameters  play equivalent roles.

It is important to emphasize that a root of   \([\omega -E -\Lambda(\omega)]\)
is not a maximum of the spectral function if  this root is within the gapless
region of the density of states. In fact, given a function \(g(x)=x/[f^{2}(x)+x^{2}]\) where \(f(x)\)
is a function with a local peak at \(x=x_{0}\), then the corresponding peak in \(g(x)\)
will be at \(x'_{0}=x_{0}\{1+1/[f'(x_{0})]^{2}\}^{-1/2}\) and therefore \(x \approx x_0\) if \(f'
\gg1\) or \(x \approx 0\) if \(f' \ll 1\). Furthermore, if the root is near the Fermi level,
the respective peak in the spectral function becomes
very narrow, but the total spectral weight associated with the peak becomes very low.
This can be easily concluded since  the spectral function [Eq.~(\ref{eq:spectralcont})]
for \(\gamma=1\) can be written in the neighborhood
of the roots $\omega_i$\ (if $\vert \omega_i \vert <D_1$) as $c\vert \omega \vert
\cal F(\omega),$ where $\cal F (\omega)$ is a normalized Lorentzian with
center at $\omega_{i}/[1+\kappa^2]$ and half-width
$\vert \kappa/(1+\kappa^2)  \cdot \omega_i \vert,$ where $\kappa=\pi \alpha V^{2}/\vert 1 -\Lambda'(\omega_i)\vert$ and
$
        c=\pi \vert 1- \Lambda' (\omega_i )\vert^{-1} \vert \kappa/(1+\kappa^2) \vert.
$
For   small $\vert\omega_i\vert$, \(\Lambda'(\omega_i)\) becomes large and \(\kappa\)
becomes small, and therefore one  has a narrow peak with a low spectral weight.
The linear factor leads to a shift of the maximum of the Lorentzian away from
zero frequency to $\omega_{i}/[1+\kappa^2]^{1/2}$.
\section{The Kondo resonance ($U = \infty$)}

In the $U \neq 0$ case, the equation of motion for the last term in Eq.~(\ref{eq:eom}) has to
be considered. Adopting the Appelbaum and Penn approximation,\cite{Appelbaum1969} one finds the
following expression for the impurity Green's function for $U=\infty$,
\begin{multline}
        \left(\omega-E_0-\Sigma_0-\Sigma_1-\Sigma_2\right)  \la\la
        d_{\s};d_{\s}^{+}\ra\ra_\omega =  \\
        1-\langle n_{d\bs} \rangle -V
        \sum_{k}\frac{\langle d_{\s}^{+}c_{k\bs}\rangle }{\omega-\varepsilon_{k}}
        , \label{eq:greensint}
\end{multline}
where
\begin{eqnarray}
        \Sigma_1 &=&
        V^2 \sum_{k,k'}
        \frac{\langle c_{k'\bs}^{+}c_{k\bs}\rangle }{\omega-\varepsilon_{k}} ,\\
        \Sigma_2 &=& V^3\sum_{k,k'}
        \frac{\langle d_{\bs}^{+}c_{k\bs}\rangle }{(\omega-\varepsilon_{k})(\omega-\varepsilon_{k'})},
\end{eqnarray}
and \(\Sigma_0\) is given by Eq.~(\ref{eq:sigma0}). In order to describe qualitatively the Kondo resonance, one  has to understand the temperature behavior
of the low order self-energy   $\Sigma_1$. For sufficiently small $V$,
one can neglect \(\Sigma_2\) and the Kondo resonance will be generated by \(\Sigma_1\), which is
approximately given by
\begin{eqnarray}
        \frac{\Sigma_1 (\omega,T)}{V^2}
        & \approx &
        - i  \pi \rho (\omega) f(\omega) \nonumber \\
        &+&  \int d\omega' \rho (\omega') f(\omega') P\left(\frac{1}{\omega-\omega'
        } \right).
        \label{eq:sigma1}
\end{eqnarray}
The Fermi-Dirac distribution function \( f(x)\) sets  a high energy cutoff in the previous integral which, for \(\gamma=1\), is approximately given by
\begin{multline}
        \text{Re}\left[\frac{\Sigma_1 (\omega,T)}{V^2}\right]
        \approx \rho_0 \ln \left\vert \frac{ D+ \omega  }{ D_{1}+ \omega  } \right\vert
        + \alpha (D_1-k_{B}T) \\ + \alpha \omega
        \ln\frac{ \sqrt{\omega^2+(a k_{B}T)^{2}}}{\vert D_1 + \omega \vert},
\end{multline}
for $k_{B}T<D_1$ and \(a\sim \pi\). We would like to note that in our numerical solution of
Eq.~(\ref{eq:greensint}), the self-energy \(\Sigma_2\) is taken into account.

\begin{figure}[tp]
        \includegraphics[width=0.45 \textwidth,clip]{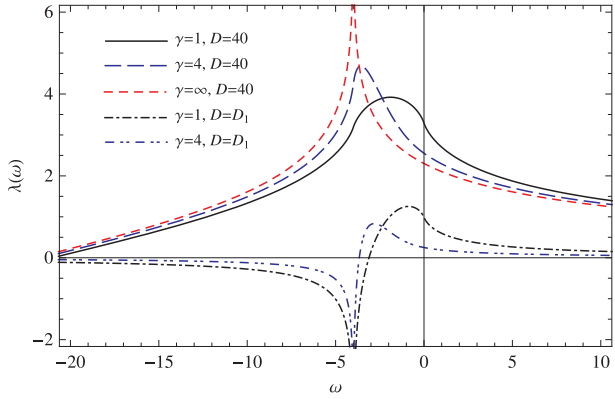}
        \caption{\label{fig:selfenergy1}
        The real part of the zero temperature self-energy \(\lambda_1(\omega)=\text{Re}\,[\Sigma_1(\omega,0)]/V^{2} \) for
         several values of \(\gamma\). In the case of
        \(\gamma=\infty\), there is a gap,  \(N(\epsilon)=0\) at \(\vert \omega
        \vert <D_1\), which leads to a logarithmic divergence of
        the real part of the self-energy at the gap edge. Other parameters: $ \rho_0=1$, $D_1=4$,
and $D=40$ or \(D=D_1\).   }
\end{figure}
The previous expression should be compared with the Lacroix result for
a constant density of states,\cite{LACROIX1981}
\begin{equation}
        \text{Re}\left[\frac{\Sigma_1 (\omega,T)}{V^2}\right]
        \sim -\rho_0\ln  \frac{\sqrt{\omega^2+(\pi k_{B}T)^{2}}}{D}.
\end{equation}
The most relevant difference between the previous two expressions is the temperature behavior.
For a constant density of states, $\text{Re}(\Sigma_1)$ diverges at zero temperature as $\omega
\rightarrow 0$.  For the  gapless density of states, as the temperature goes to zero, $\text{Re}(\Sigma_1)$
remains finite at the Fermi energy (as shown in Fig.~\ref{fig:selfenergy1}) with a peak in the energy range
\(-D_1< \omega <0\). Note that for a strong depletion of the density of states  this peak grows and approaches
the gapless region edge, converging therefore to the logarithmic  divergence of the real part of the
self-energy of an equivalent
system with a $2D_1$ gap. In Fig.~\ref{fig:selfenergy1}, the self-energy for a linear and the quartic  density
of states with \(D=D_{1}\) is also displayed.  The self-energy of the latter is strongly
reduced in the \(-D_{1}<\omega <0\) region due to the the absence of the log-like contribution which
resulted from the \(-D<\omega <-D_{1}\) constant term in the density of states.
\subsection{The critical hybridization}

The critical temperature for the Kondo resonance in the equation of motion approach \cite{LACROIX1981}
is associated with  a temperature  below which a new solution of \(\text{Re} [\la\la
        d_{\s};d_{\s}^{+}\ra\ra_\omega^{-1}]=0\) appears. This solution is due to an additional intersection between the line \(\omega -E_0\) and
the real part of the self-energy \(\text{Re}(\Sigma_1) \). In the case of a constant density
of states, this solution occurs near \(\omega =0\). Since the maximum value of the self-energy for the constant density
of states ($\gamma=0$) is
\begin{equation}
        \text{Re}\left[\frac{\Sigma_1 (0,T) }{V^2}\right]\sim -\rho_0\ln \left(  \frac{\pi k_{B}T}{D}\right)
\end{equation}
and decreases with increasing temperature, above a critical temperature one
loses the additional intersection and the Kondo peak disappears. The condition
$
        E_{0}/V^{2}\sim \rho_0\ln (k_{B}T_{c}/D)
$
determines the Kondo temperature. Note that this condition has a solution  for any value of \(V\), however small it may be, due to the logarithmic
dependence on temperature.

\begin{figure}[tp]
      \includegraphics[width=0.45 \textwidth,clip]{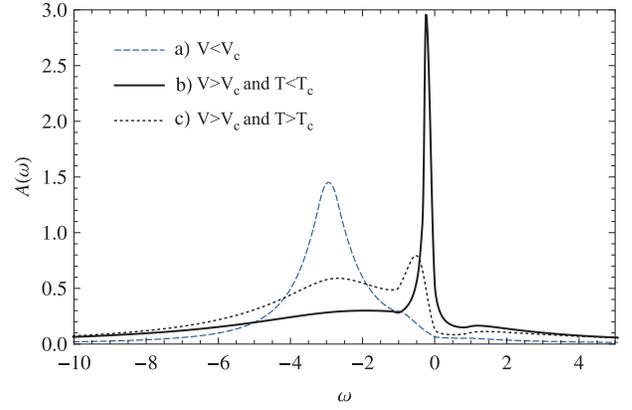}
      \caption{\label{fig:spectralintV}  The spectral function \(A(\omega)\),  for \(U=\infty\), \(\gamma=1\), and several values of the hybridization \(V\).  Curve a) displays the typical \(V<V_{c} \) spectral function
characterized by the absence of a Kondo resonance. The Kondo  peak  appears as \(V\) is increased above a critical value \(V_{c}\) [curve b)] and disappears
above the Kondo temperature \(T_{c}\) [curve c)]. Other parameters:  \(E_{0}=-3.2\),
        $\rho_0=10$, $D_1=1$, $D=20$ and a) $V=0.12$, $k_{B}T=0.05$; b) $V=0.24$, $k_{B}T=0.05$;
        c) $V=0.24$, $k_{B}T=10$. }
\end{figure}
In the case of the gapless density of  states (with $\gamma \geq 1$), the same reasoning can be
followed in order  to obtain the critical temperature. However, the real part of the self-energy \(\Sigma_1 \) does not diverge
at zero temperature for any frequency (excluding the band edges). Therefore,
for a sufficiently small hybridization, the Kondo peak does not occur. In the case of
\(\rho(\omega)=\alpha \vert \omega \vert^\gamma  \Theta (D_1 - \vert \omega \vert) \), the
real part of the  self-energy \(\)  for large \(\gamma\) can be obtained expanding
the density of states at \(\omega=-D_1\) and is given approximately by
\begin{eqnarray}
        \text{Re}\left[\frac{\Sigma_1 (\omega,0)}{V^2}\right]
         \approx    \alpha D_1^\gamma
        & + & \alpha D_1^\gamma\left[1- \gamma
        \left( 1+ \frac{\omega}{D_1} \right) \right] \nonumber
        \\ & \times & \ln \left\vert 1-
        \frac{ 1/\gamma  }{1+ \frac{\omega}{D_{1}}  } \right\vert
\end{eqnarray}
This function has a maximum at \(\omega_{max}\) given by  \(\omega_{max} +D_1 \sim 3D_{1}/4\gamma\)
and  approximately  of the value of \(\rho_0 \).
The constant behavior of the density of states in the energy interval  \(-D<\omega <-D_{1}\) leads to
an additional \(\rho_0 V^{2}\ln \left\vert ( D+ \omega  )/( D_{1}+ \omega  ) \right\vert \)
term in the real part of the self-energy,
$\text{Re}\left[\Sigma_1 (\omega,0)\right]$. For \(D \gg D_1\), the maximum of the real part of
the  self-energy is shifted to $\omega_{max} +D_1 \sim D_{1}/2\gamma$ and its value is approximately
$
         \rho_0 V^2\ln \left(  2\gamma
        { D  }/{ D_{1}   } \right)
$
which is the same as the former maximum value except for an enhancement
factor \(\ln \left(  2\gamma { D  }/{ D_{1}   } \right)\).
Note that this enhancement reflects the contribution of the constant density
of states region. In fact, for large \(\gamma\), the behavior becomes similar to a gapped system.
In Fig.~\ref{fig:selfenergy1}, the difference between the maximum values of  \(\text{Re}\left[\Sigma_1 (\omega,0)\right]\) for $D\gg D_1$ and $D=D_1$ is clearly observed as well as the
shift of the maximum position (with increasing $\gamma$) towards the negative edge
of the gapless region.
\begin{figure}[tp]
      \includegraphics[width=0.45\textwidth,clip]{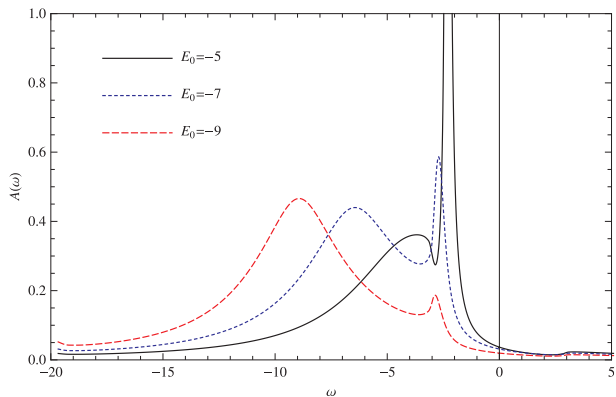}
      \caption{\label{fig:spectralE0}
      The spectral function \(A(\omega)\)  for \(U=\infty\), \(\gamma=8\), and several values
of the impurity energy, $E_{0}=-5$,
        $-7$ and $-9$. One can see that when  \(E_0\) increases and approaches the edge of the
depletion region (\(\omega=-3\), on this figure), the Kondo peak grows within the depletion
region much below the Fermi energy.  Other parameters:
$\rho_0=10$, $D_1=3$, $D=20$, $V=0.2$, and $k_{B}T=0.05$.  }
\end{figure}

The critical hybridization is the minimum \(V\) which is necessary  to have
an intersection between the line $\omega -E_0$ and \(\text{Re}\left[\Sigma_1 (\omega,0)\right]\)
at zero temperature. For \(E_{0}\lesssim \ -D_1\),
one has to take into account that the  maximum occurs at \(\omega_{max} \)
and the condition for the critical hybridization becomes
for $\gamma \gtrsim 1$
\begin{equation}
        \frac{\omega_{max}-E_{0} }{V^{2}}\approx \left\{
        \begin{array}{lc}
                \alpha D_1^\gamma \ln \left(  2\gamma { D  }/{ D_{1}   }  \right), & D \gg D_1, \\
                \alpha D_1^\gamma, & D=D_1.
        \end{array} \right.
        \label{eq:criticalcond}
\end{equation}
It is necessary to note that the density of states used in the renormalization group approach
\cite{Withoff1990} corresponds to  $D=D_1$. For not so large $\gamma $, this density of states
is reasonable, but in the $\gamma \rightarrow \infty$ limit, this density of states becomes unrealistic
since the  electron states  are  concentrated in a narrow region at the band edges. Equation~(\ref{eq:criticalcond}) should be compared with the expression for the critical exchange coupling
\(J_{c}\) of the spin-1/2 Kondo model with a gapless conduction band, \(J_{c} \approx
\gamma/\alpha D_1^\gamma =\gamma/ \rho_0\).\cite{Withoff1990} The Kondo  model can be obtained from the Anderson model
within perturbation theory \cite{Hewson1993}
leading to an exchange  coupling
$
        J_{\mathbf{k}\mathbf{k'}} \sim V^2/(\varepsilon_\mathbf{k} - E_0)
$
which is antiferromagnetic for energies \(\varepsilon_\mathbf{k} \sim 0\).
The resonant scattering of the conduction electrons in that energy region
leads to the usual Kondo behavior with
$
        J \sim -V^2/ E_0
$.
In fact, the Kondo limit corresponds to taking the limits
\(V \rightarrow \infty\) and \(E_0 \rightarrow -\infty\) (as well as \(U\rightarrow \infty\))
in the Anderson model.
That is why the  correction due to \(\omega_{max}\) is not observed  in  the renormalization
studies of the Kondo model \cite{Withoff1990,Sengupta2008} and has been also ignored in renormalization
studies of the Anderson model.\cite{Gonzalez-Buxton1998}

For a gapped conduction band (\(\gamma= \infty\)) with a $2D_1$ gap, it is straightforward to conclude that the exchange coupling becomes
$
        J \sim V^2/ (D_1-E_0)
$.
In the case of a gapless conduction band, one also has to take into account
the depletion of states near the Fermi level and a correction in the previous expression should
appear. In fact, a continuous evolution between the two cases should occur as $\gamma$ is
varied. The real part of the self-energy, Eq.~(\ref{eq:sigma1}),  and in particular its maximum,
reflect  this evolution. Consequently, in the gapless case one has
$ J_{} \sim V^2/(\omega_{max}-E_{0})$ and  \(J_c^{-1}\)
is given by Eq.~(\ref{eq:criticalcond}) which  shows\ two different scenarios for the gapless energy region in what concerns the Kondo
resonance. For a pure power-law density of states
with $\gamma \gtrsim 1$, the equivalent  critical
coupling will be \(J_{c} \sim 1/\alpha D_1^\gamma =1/\rho_0 \) and therefore is reduced in comparison with the renormalization group result \cite{Withoff1990} in the spin-1/2 Kondo  model by a factor \(\gamma\). For \(D\gg D_1\),  the gapless energy region provides a cutoff
to the  log-like behavior of the real part of the self-energy for \(\gamma=\infty\) and
\begin{equation}
        J_{c}^{-1} \sim \alpha D_1^\gamma \ln (2\gamma D/D_1)= \rho_0 \ln (2\gamma D/D_1)
        \label{eq:jc}
\end{equation}
and
therefore an important logarithmic  factor
is present. Note that according to Eq.~(\ref{eq:jc}) as \(\gamma\) becomes larger, the critical coupling goes to zero, contrary to the renormalization group result,  \(J_{c} \approx
\gamma/\rho_0 \). We suggest that this discrepancy is due to the  unrealistic behavior of the
density of states used in the renormalization group approach, see the discussion above. The behavior described by Eq.~(\ref{eq:jc}) reflects  the existence of a Kondo-like  peak in the impurity spectral  function in the strong depletion limit, \(\gamma \gg 1\) (see Fig.~\ref{fig:spectralE0}).
This phenomenon can be related to the Shiba intragap states that occur near the band edges in s-wave
superconductors.\cite{Balatsky2006}

\begin{figure}[tp]
      \includegraphics[width=0.45\textwidth,clip]{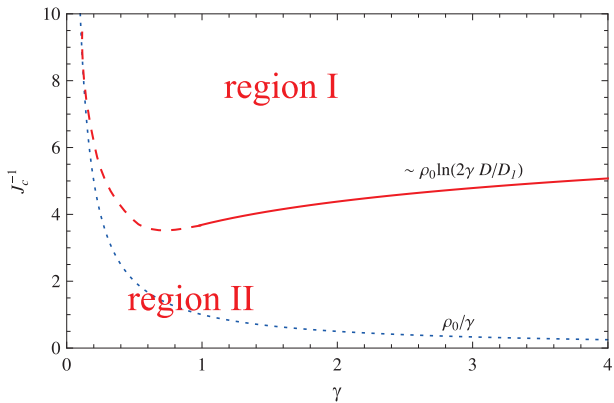}
      \caption{\label{fig:jc}  Phase diagram of the gapless Anderson model in
the $\gamma-J_c^{-1}$ plane. The results presented
in this paper are described by the solid red  line.  Above the solid red line (region I), i.e., in the region with a sufficiently small exchange coupling  (or equivalently, a small hybridization
parameter \(V\)), the Kondo-like peak is not formed. Strong electron-electron correlations
lead to a strong peak  in the impurity spectral function in region II, i.e., the region of
a strong exchange coupling (\(J>J_{c}\)) and a strong hybridization (\(V>V_{_{c}}\)). The
 blue  dotted curve represents the renormalization group result of Withoff and Fradkin \cite{Withoff1990}. }
\end{figure}

In Fig.~\ref{fig:spectralintV}, the spectral function for \(U=\infty\) is displayed  for several values of \(V\) and for temperatures
above and below the Kondo temperature. These results were obtained numerically from Eq.~(\ref{eq:greensint})
taking into account all  self-energy terms. Curve a) displays the typical behavior of the impurity
spectral function when \((\omega_{max}-E_{0} )/V^{2}\) is above the critical value
and no Kondo resonance is  observed (even at zero temperature).
In curve b) of Fig.~\ref{fig:spectralintV}, a Kondo peak is present since \((\omega_{max}-E_{0} )/V^{2}\) is below the critical value. Note that this Kondo peak does not occur at \(\omega=0\) reflecting the
fact that the maximum of the real part of the self-energy $\Sigma_1$ occurs at \(\omega_{max}\) in contrast
to the usual Kondo behavior.
A broader and lower Lorentzian profile
is also observed in curve b) in Fig.~\ref{fig:spectralintV}
reflecting a larger value of
the hybridization.
Increasing temperature above the Kondo temperature, the narrow
Kondo
peak disappears but a small broad peak may remain due to the depletion of the density of states
at the Fermi level as explained in the discussion of the \(U=0\) case. This is the situation
displayed in curve c) of Fig.~\ref{fig:spectralintV}.

For large $\gamma$,  the  Kondo peak emerges within the
depletion region near its  negative edge as shown in Fig.~\ref{fig:spectralE0}. Lowering
the impurity energy level leads to the disappearance of the Kondo peak in agreement with the
condition given by Eq.~(\ref{eq:criticalcond}). Note that the Lorentzian profiles have similar heights and widths in  Fig.~\ref{fig:spectralE0} since the hybridization energy
is the same for all curves.

The critical coupling relation [Eq.~(\ref{eq:jc})] leads to the qualitative phase diagram of
Fig.~\ref{fig:jc}. The renormalization group result is also displayed
in this figure.  A considerably larger region (region II) of resonant Kondo-like correlations
induced by the strong on-site Coulomb repulsion is present in comparison with the renormalization
group result (dotted blue curve in Fig.~\ref{fig:jc}). This deviation becomes larger as $\gamma$ grows. Note that, in this paper,  we have not addressed the $\gamma<1$ case. However, it is reasonable that as $\gamma$ goes to
zero, the constant density of states behavior is recovered. Our belief  is that the renormalization
group  approach  describes  correctly the  $\gamma \rightarrow 0$ limit and therefore,  \(J_{c}^{-1}  \propto 1/\gamma \) for $\gamma \ll 1$.  The dashed part of the red curve in
Fig.~\ref{fig:jc} reflects this assumption.

\section{Conclusion}

In conclusion, we have determined the critical hybridization for the Kondo effect
in the gapless Anderson model. Previous works have used  renormalization group approaches to the
Kondo effect in gapless systems and  considered a purely power-law density of states.
Here we have shown that if one considers a more realistic density of states, the
critical coupling expression is  reduced by a logarithmic factor that reflects the fact that the
Kondo effect is dominated by the contributions of the density of states away from the gapless
energy
region. We showed that in a gapless system, strong electron-electron correlations between conduction
and impurity electrons result in a sharp Kondo-like peak if the hybridization is larger than
a critical value. The critical hybridization depends on the behavior of the density  of states
in the depletion region and on the energy of the impurity level. The  low energy critical exchange coupling \(J_{c}\)  for the emergence of the Kondo
resonance reflects the depletion
of states at the Fermi energy and   therefore is determined by the contribution of the states near the gapless region edge. We have demonstrated that in gapless systems, due to the strong on-site Coulomb repulsion, a sharp Kondo peak emerges  much below the  Fermi energy   if the
density of states is strongly depleted. This Kondo peak has a   strong dependence on temperature. Although the results presented above rely in a specific form of the density of states, we believe that any system with a well behaved density of states (except for its gapless behavior at the Fermi energy) should display
similar behavior.

\bibliography{kondo1}

\end{document}